\newcommand{\ignore}[1]{}
\documentclass[11pt]{article}
\newcommand\be{\begin{equation}}
\newcommand\ee{\end{equation}}
\newcommand\bea{\begin{eqnarray}}
\newcommand\eea{\end{eqnarray}}
\newcommand\half{{\textstyle{1\over2}}}

\def\bmat{      \left |  \begin{array}{cc} }
\def\emat{ \end{array} \right |    }

\usepackage{graphicx}
\def\be{\begin{equation}}
\def\ee{\end{equation}}
\def\bea{\begin{eqnarray}}
\def\eea{\end{eqnarray}}

\begin{document}
\begin{titlepage}
\noindent BROWN-HET-1100:TA552  \hfill Oct. 17, 1997 
\begin{center}

{\Large\bf Flavoring of Pomeron and Diffractive Production at Tevatron Energies}

\vspace{3cm}
{\bf Chung-I Tan$^{(1)}$}\\
\end{center}
\vspace{1.0cm}
\begin{flushleft}
{}~~$^{(1)}$Department of Physics, Brown University, Providence, RI 02912,
USA\\
\end{flushleft}
\vspace{3cm}


\abstract{The most important consequence of Pomeron
being a pole is the factorization property.
 However, due to 
Pomeron intercept being greater than 1,  the extrapolated 
single diffraction dissociation cross section based on a classical triple-Pomeron formula is too
large leading to a potential unitarity violation at Tevatron energies.  
 It is
our desire here to point out that  the ``flavoring" of  Pomeron plays the
dominant role in resolving this apparent ``paradox".   }

\vfill
\noindent Talk presented at VIIth Blois Workshop on Elastic and Diffractive
Scattering, Seoul, Korea, (1997).
\end{titlepage}

\section{Introduction}
One of the more interesting developments  from
recent collider experiments 
 is the finding that  hadronic  total cross sections as well as  elastic cross sections in the
near-forward limit can be  described by the exchange of a ``soft Pomeron"
pole,~\cite{Tan1} {\it i.e.},
 the absorptive part of the elastic amplitudes can be approximated by 
${Im}\> T_{a,b}(s,t)\simeq \beta_a(t) s^{\alpha_{\cal P}(t)}\beta_b(t).$
 The Pomeron trajectory has two important features. First, its
zero-energy intercept is greater than one,
$\alpha_{\cal P}(0)\equiv 1+
\epsilon$,
$\epsilon\simeq 0.08\sim 0.12$, leading to rising
$\sigma^{tot}(s)$. Second, its Regge slope is approximately
$\alpha_{\cal P}'\simeq 0.25\sim 0.3$ $ GeV^{-2}$, leading to the
observed shrinkage effect for elastic peaks.  The most important
consequence of Pomeron being a pole is factorization. For a singly
diffractive dissociation process, factorization leads to a ``classical
triple-Pomeron" formula,~\cite{Classical} $ {d\sigma \over
dtd\xi}\rightarrow {d\sigma^{classical} \over dtd\xi}\equiv F_{{\cal
P}/a}^{cl} (\xi, t) \sigma_{{\cal P}b}^{cl} (M^2,t), $ where $M^2$ is
the missing mass variable and $\xi\equiv M^2/s$.  The first term, $
F_{{\cal P}/a}^{cl} (\xi, t)$, is the so-called ``Pomeron flux", and
the second term is the ``Pomeron-particle" total cross section.  With
$\epsilon\sim 0.1$, it has been observed~\cite{Dino1} that the
extrapolated $p\bar p$ single diffraction dissociation cross section,
$\sigma^{sd}$, based on the standard triple-Pomeron formula is too
large at Tevatron energies by as much as a factor of $5\sim 10$ and it
could become larger than the total cross section.

Let us denote the singly diffractive cross section as a product of a
``renormalization" factor and the classical formula,
\be
{d\sigma \over dtd\xi}=Z(\xi,t;s) {d\sigma^{classical} \over dtd\xi}.
\label{eq:TotalRenormalization}
\ee
It was argued by K. Goulianos in Ref. 3 that agreement with data could
be achieved by having an energy-dependent suppression factor,
$Z(\xi,t; s)\rightarrow Z_{G}(s)\equiv N(s)^{-1}\leq 1$, so that the
new ``Pomeron flux", {$F_N(s, \xi, t) \equiv N(s)^{-1}F_{{\cal
P}/p}^{cl} (\xi, t)$}, is normalized to unity for {$ s\geq \bar s$},{
$\sqrt{\bar s}\simeq 22\> GeV$}.  An alternative suggestion has been
made recently by P.  Schlein,~\cite{Schlein1} where $Z\rightarrow
Z_{S}(\xi)$.

 In view of the factorization property for total and elastic cross
 sections, the ``flux renormalization" procedure appears paradoxical.
 We shall refer to this as ``{\bf Dino's paradox}''.  {Finding a
 resolution that is consistent with Pomeron pole dominance for elastic
 and total cross sections at Tevatron energies will be the main focus
 of this study.}  In particular, we want to maintain the following
 factorization property, ${ d\sigma \over dtd\xi}\rightarrow \sum_k
 D_k(\xi,t)\Sigma_k(M^2), $ when $\xi^{-1}$ and $M^2$ become
 large.~\cite{Tan1}

It is generally expected that  the
resolution to the paradox should lie in a proper implementation of screening corrections to the
classical triple-Pomeron formula.  
 Our treatment  lies in a proper implementation of final-state screening correction, (or
{\bf final-state unitarization}), with ``{\bf flavoring}'' for Pomeron as the primary dynamical
mechanism for setting the relevant energy scale. In our treatment, initial-state screening remains
unimportant, consistent with the pole dominance picture for elastic and total cross section
hypothesis at Tevatron energies.  
In fact, we shall concentrate in the present discussion only on the flavoring
whereas the treatment of final-state screening can be found in the Ref. 1
and the effect turns out to be small.

\section{Dynamics for Soft Pomeron and Flavoring}

Although we have shown in Ref. 1 that final-state screening would automatically avoid unitarity
violation, the primary source of high energy suppression actually comes from a proper treatment of
{\bf scale-dependence} for Pomeron couplings.  Consider for
the moment the following scenario where one has two different fits to hadronic total cross sections:
\begin{itemize}
\item
(a) ``High energy fit": $\sigma_{ab}(Y)\simeq \beta_a\>\beta_b \>e^{\epsilon\> Y}$ \hskip50pt for
\hskip50pt
$Y>>y_f$,
\item
(b) ``Low energy fit": $\sigma_{ab}(Y)\simeq \beta^{low}_a\>\beta_b^{low} $ \hskip50pt for
\hskip40pt $0<Y<<y_f$.
\end{itemize}
That is, we envisage a situation where the ``effective Pomeron intercept", $\epsilon_{eff}$, 
increases from $0$ to $\epsilon\sim 0.1$ as one moves up in energies.   In order to have  a
smooth interpolation between these two fits, one can obtain the following order of magnitude estimate
$
\beta_p\simeq e^{-{\epsilon\> y_f\over 2}} \beta_p^{low}.
$
 Modern parametrization for Pomeron residues typically
leads to values of the order $(\beta_p)^2\simeq  14\sim 17 $ mb. However, before the advent of the
notion of a Pomeron with an intercept greater than 1, a typical parametrization would have a value
$(\beta^{low}_p)^2\simeq 35\sim 40$ mb, accounting for a near constant Pomeron contribution at low
energies. This leads to an estimate of $y_f\sim 8$, corresponding to $ \sqrt s \sim 50$ GeV.
This is precisely the energy scale where a rising total cross section first becomes noticeable.

The scenario just described has been referred to as ``flavoring", the notion that the
underlying effective degrees of freedom for Pomeron will increase as one moves to higher
energies,~\cite{Flavoring1}   and it has provided a dynamical basis
for understanding the value of Pomeron intercept in a non-perturbative QCD
setting.~\cite{Flavoring4} In this scheme, both the Pomeron intercept and the
Pomeron residues are {\bf scale-dependent}. We shall briefly review this mechanism and introduce a
scale-dependent formalism where  the entire flavoring effect can be absorbed into a flavoring factor,
$R(y)$, associated with each Pomeron propagator. 

\subsection{Bare Pomeron in Non-Perturbative QCD}

In a non-perturbative QCD setting, the Pomeron intercept is
directly related to the strength of the short-range order component of inelastic production and this
can best be understood in a large-$N$ expansion.~\cite{LeeVeneziano} In such a scheme,
particle production mostly involves emitting ``low-mass pions", and  the basic energy scale of
interactions is that of ordinary vector mesons, of the order of 
$1$ GeV.  In a one-dimensional multiperipheral realization for the ``planar
component" of the large-$N$ QCD expansion, the high energy behavior of a $n$-particle total cross
section is primarily controlled by its longitudinal phase space,
$\sigma_n\simeq (g^4N^2/{(n-2)!})(g^2N\log s)^{n-2} s^{J_{eff}-1}.$
Since there are only Reggeons at the planar diagram level, one has  $J_{eff}=2\alpha_R-1$ and, after
 summing over $n$,  one  arrives at Regge behavior for the planar component of $\sigma^{tot}$ where 
\be
\alpha_R=(2\alpha_R-1)+g^2N.
\label{eq:Planar}
\ee
At next level of cylinder topology,  the contribution to partial cross section increases due to its topological twists,
$\sigma_n\simeq {( g^4/ {(n-2)!})} 2^{n-2}(g^2N\log s)^{n-2} s^{J_{eff}-1},$
and, upon summing over $n$,  one arrives at a total cross section governed by a Pomeron exhange, 
$
\sigma_0^{tot}(Y)=g^4e^{\alpha_{\cal P} Y},
$
where the  Pomeron interecept is 
\be
\alpha_{\cal P}=(2\alpha_R-1)+2g^2N.
\label{eq:Cylinder}
\ee
Combining Eq. ({\ref{eq:Planar}) and Eq. (\ref{eq:Cylinder}), we arrive at an
amazing ``bootstrap" result,
$
\alpha_{\cal P}\simeq 1.
$

In a non-perturbative QCD setting, having a Pomeron intercept near 1
therefore depends crucially on the topological structure of large-$N$
non-Abelian gauge theories.~\cite{LeeVeneziano} In this picture, one
has $\alpha_R\simeq .5\sim .7$ and $g^2N\simeq .3\sim .5 $.  With
$\alpha'\simeq 1$ $GeV^{-2}$, one can also directly relate $\alpha_R$
to the average mass of typical vector mesons.  Since vector meson
masses are controlled by constituent mass for light quarks, and since
constituent quark mass is a consequence of chiral symmetry breaking,
the Pomeron and the Reggeon intercepts are directly related to
fundamental issues in non-perturbative QCD.

Finally we note that, in a Regge expansion, the relative importance of
secondary trajectories to the Pomeron is controlled by the ratio
\begin{equation}
e^{\alpha_{R}\> y}/e^{\alpha_{\cal P}\> y}= e^{-(\alpha_{\cal
P}-\alpha_{\cal R})\> y}. 
\end{equation}
It follows that there exists a natural
scale in rapidity, $y_r$, $(\alpha_{\cal P}-\alpha_R)^{-1}<y_r \simeq
3\sim 5$.  The importance of this scale $y_r$ is of course well known:
When using a Regge expansion for total and two-boby cross sections,
secondary trajactory contributions become important and must be
included whenever rapidity separations are below $3\sim 5$ units. This
scale of course is also important for the triple-Regge region: There
are two relevant rapidity regions: one associated with the ``rapdity
gap", $y\equiv \log \xi^{-1}$, and the other for the missing mass,
$y_m\equiv \log M^2$.

\subsection{Flavoring of Bare Pomeron}

We have proposed sometime ago  that ``baryon pair" and ``heavy flavor" production provides an
additional energy scale, $s_f=e^{y_f}$,  for soft Pomeron
dynamics,  and this effect
can be  responsible for the perturbative increase of the Pomeron intercept to be greater than unity, 
$\alpha_{\cal P}(0)\sim
1+\epsilon,\>\>\epsilon>0$. One must bear this additional energy scale in mind in working with a soft
Pomeron.~\cite{Flavoring4}
That is, to fully justify using a Pomeron with an intercept $\alpha_{\cal P}(0)>1$, one must restrict
oneself to energies
$s>s_f$ where heavy flavor production is no longer suppressed. Conversely, to extrapolate Pomeron
exchange to low energies below $s_f$, a lowered ``effective trajectory" must be
used. This feature of course is
unimportant for total and elastic cross sections at Tevatron energies. However, it is important for
diffractive production since both $\xi^{-1}$ and $M^2$ will sweep right through this energy  scale at
Tevatron energies.

Flavoring becomes important whenever  there is a further inclusion of effective degrees of freedom
than that associated with light quarks. This can again be illustrated by a simple one-dimensional
multiperipheral model. In addition to what is already contained in the Lee-Veneziano
model, suppose that new particles can also be produced in a multiperipheral
chain. Concentrating on the cylinder level, the partial  cross sections will be  labelled by
two indices, 
\be
\sigma_{p,q}\simeq  (g^4/ p!q!) 2^{p+q}(g^2N\log s)^{p} (g^2_{f}N\log
s)^{q}s^{J_{eff}-1},
\label{eq:FlavoringTotalCrossSection}
\ee
where $q$ denotes the number of clusters of new particles produced. Upon summing over $p$ and
$q$, we obtain a ``renormalized" Pomeron trajectory
\be 
\alpha_{\cal P}=\alpha^{old}_{\cal P}+ \epsilon,
\label{eq:NewPomeron}
\ee
where $\alpha^{old}_{\cal P}\simeq 1$ and $\epsilon\simeq  2g^2_{f}N$. That is, in a non-perturbative
QCD setting, the effective intercept of Pomeron is a dynamical quantity,  reflecting the effective
degrees of freedom involved in near-forward particle
production.\cite{Flavoring4}

If  the new degree of freedom involves
particle production with high mass, the longitudinal phase space factor, instead of $({\log
s})^q$, must be modified. Consider the situation of producing one $N\bar N$ bound state together
with pions, {\it i.e.}, $p$ arbitrary and  $q=1$ in Eq. (\ref{eq:FlavoringTotalCrossSection}). 
Instead of
$(\log s)^{p+1}$,  each factor  should be replaced by
$(\log(s/m^2_{eff}))^{p+1}$, where $m_{eff}$ is an effective  mass for the $N\bar N$ cluster.
In terms of rapidity, the longitudinal phase space factor becomes
$(Y-\delta)^{p+1}$,  where
$\delta$ can be thought of as a one-dimensional ``excluded volume" effect. 
For heavy particle production, there will be an energy range over which keeping up to $q=1$
remains a valid approximation. Upon summing over $p$, one finds that the additional contribution
to the total cross section due to  the production  of one heavy-particle cluster
is~\cite{Flavoring1} 
$
\sigma^{tot}_{q=1}\sim \sigma_0^{total}(Y-\delta)(2g^2_fN)\log
(Y-\delta)\theta(Y-\delta),
$
where $\alpha_{\cal P}^{old}\simeq 1$. 
Note the effective longitudinal phase space ``threshold
factor",
$\theta(Y-\delta)$, and, initially, this term  represents a small perturbation  to the total
cross section obtained previously, (corresponding to
$q=0$ in Eq. (\ref{eq:FlavoringTotalCrossSection})), $\sigma_0^{total}$. 
Over a rapidity range, $[\delta, \delta+\delta_f]$, where $\delta_f$ is the average rapidity
required for producing another heavy-mass cluster, this is  the only term needed for incorporating
this new degree of freedom. As one moves to higher energies,
``longitudinal phase space suppression" becomes less important and more and more heavy particle
clusters will be produced. Upon summing over
$q$, we would obtain a new total cross section, described by  a  renormalized Pomeron,  with a new
intercept given by Eq. (\ref{eq:NewPomeron}).

We  assume that, at Tevatron, the energy is high enough so that this kind of ``threshold"
effects is no longer important.  How low an energy do we have to go before one encounter these
effects? Let us try to answer this question by starting out from low energies. As we have stated
earlier, for
$Y> 3\sim 5$, secondary trajectories become unimportant and using  a Pomeron with
$\alpha_{\cal}\simeq 1$ becomes a useful approximation. However, as new flavor production becomes
effective, the Pomeron trajectory will have to be renormalized. We can estimate for the relevant
rapidity range when this becomes important as follows:  $y_f >  2
\delta_{0}+
<q>_{min} \delta_f$. The first factor
$\delta_{0}$ is associated with leading particle effect, i.e., for proton, this is primariy due
to pion exchange.
$\delta_f$ is the minimum gap associated with one heavy-mass cluster  production, {\it  e.g.},
nucleon-antinuceon pair production. We estimate
$\delta_{0}\simeq  2$ and $
\delta_f\simeq 2\sim 3 $,  so that, with $<q>_{min}\simeq 2$,  we  expect the
relevant flavoring rapidity scale to be 
$y_f\simeq 8\sim 10$.

\subsection {Effective intercept  and Scale-Dependent  Treatment}

In order to be able to extend  a Pomeron repesentation below the rapidity scale $y\sim y_f$, we
propose   the following {\bf scale-dependent}  scheme where we introduce a flavoring factor for each
Pomeron propagator.  Since each
Pomeron exchange is always associated with  energy variable
$s$, (therefore a rapidity variable
$y\equiv
\log s$), we shall  parametrize the Pomeron trajectory function as 
\be
\alpha_{eff}(t; y)\simeq 1+\epsilon_{eff}(y) +\alpha' t,
\label{eq: EffectivePomeronTrajectory}
\ee
where $\epsilon_{eff}(y)$ has the properties
\begin{itemize}
\item 
{(i)} $\epsilon_{eff}\simeq \epsilon_o\equiv \alpha^{old}_{\cal P}-1\simeq 0 $ \hskip55pt for
\hskip40pt 
$y<< y_f$,
\item 
{(ii)}  $\epsilon_{eff}\simeq \epsilon \simeq 0.1  $ \hskip100pt for \hskip40pt  $y>>y_f$.
\end{itemize}
For instance, exchanging such an effective Pomeron leads to a contribution to the elastic cross
section 
$
T_{ab}(s,t)\propto s^{1+\epsilon_{eff}(y) +\alpha' t}.
$
This representation  can now be extended down to the region $y\sim y_r$. 
We shall adopt a particularly convenient parametrization for $\epsilon_{eff}(y)$ in the next
Section when we discuss phenomenological concerns.

To complete the story, we need also to account for the scale dependence of Pomeron residues. What we
need is an ``interpolating" formula between the high energy and low energy sets. Once a choice for
$\epsilon_{eff}(y)$ has been made, it is easy to verify that a natural choice is simply
$\beta_a^{eff}(y)=\beta_ae^{[\epsilon-\epsilon_{eff}(y)]y_f}$. It follows that the total
contribution from a ``flavored" Pomeron to a Pomeron amplitude is 
 $
 T_{a,b}(y,t)= R(y)\>
T_{a,b}^{cl}(y,t),
$
where $ T_{a,b}^{cl}(y,t)\equiv \beta_a\beta_be^{(1+\epsilon+\alpha'_{\cal P}t)\> y}$ is the
amplitude according to a ``high energy" description with a fixed Pomeron intercept, and $R(y)\equiv
e^{-[\epsilon-\epsilon_{eff}(y)](y-y_f)}$ is a ``flavoring" factor. In terms of  $s=e^y$, 
$
R(s)\equiv ({s_f\over s})^{[\epsilon-\epsilon_{eff}(\log s)]}.
$

This flavoring factor
should  be consistently applied as part of each  ``Pomeron propagator". With  the  normalization 
 $R(\infty)=1$, we can therefore leave the residues alone, once they have been determined by a
``high energy" analysis.  For instance, for the single-particle gap  cross section, since there are
three Pomeron propagators, one has for the renormalization factor: 
$ Z=
R^2(y)R(y_m).
$
It is instructive to plot in Figure~\ref{fig:flavoring}  this combination  as a function of either $\xi$ or
$M^2$ for various fixed values of $Y$.

\section{ A Caricature of High Energy Diffractive Dissociation}

Both the  screening function and the flavoring function depend on the effective Pomeron intercept,
and  we shall adopt the following  simple parametrization.  The transition from
$\alpha^{old}(0)=1+\epsilon_o$ to
$\alpha^{new}(0)=1+\epsilon$ will occur over a  rapidity range,
$(y_f^{(1)}, y_f^{(2)})$. Let $ y_f\equiv \half(y_f^{(1)}+ y_f^{(2)})$ and $\lambda_f
^{-1}\equiv 
\half(y_f^{(2)}- y_f^{(1)})$. Similarly, we also define $\bar \epsilon\equiv
\half (\epsilon+\epsilon_o)$ and  $\Delta\equiv \half(\epsilon-\epsilon_o)$. A convenient
parametrization for $\epsilon_{eff}$ we shall adopt is 
$\epsilon_{eff}(y) =[\bar \epsilon +{\Delta} \tanh {{\lambda_f}
(y-\bar y_f)}].
$
The combination 
$[{\epsilon-\epsilon_{eff}(y)}]$ can be written as $(2{\bar\epsilon)\> [1+({s/
s_f})^{2\lambda_f}]^{-1}}$ where $ s_f=e^{ y_f}$. We arrive at a simple parametrization for our
flavoring function
\be
R(s)\equiv \bigl({s_f\over s}\bigr)^{(2\bar\epsilon)\> [1+({s\over \bar s_f})^{2\lambda_f}]^{-1}}.
\label{eq:FlavoringFactor2}
\ee
With
$\alpha_{\cal P}^{old}
\simeq 1$, we have $\epsilon_o\simeq 0$, $\bar \epsilon\simeq \Delta \simeq \epsilon/2$, 
and we expect that  $\lambda_f
\simeq 1\sim 2
$ and
$ y_f \simeq 8\sim 10$ are reasonable range for these
parameters.~\cite{GlobalFlavoring}

The most important new parameter we have introduced for understanding high energy diffractive
production is the flavoring scale, $s_f=e^{y_f}$. We have motivated by way of a simple model to
show that a reasonable range for this scale is $y_f\simeq 8\sim 10$. Quite independent of
our estimate, it is possible to treat our proposed resolution phenomenologically and determine this
flavoring scale from experimental data.

It should be clear that
one is  not attempting to carry out a full-blown phenomenological analysis here. To do that, one must
properly incorporate other triple-Regge contributions, {\it e.g.}, the
${\cal PPR}$-term for the low-$y_m$ region, the $\pi\pi{\cal P}$-term and/or the ${\cal RRP}$-term 
for the low-$y$ region, etc., particularly for $\sqrt s \leq \sqrt {s_f}\sim 100\> GeV$.  What we 
hope to achieve is to provide a ``caricuture" of the interesting physics involved in diffractive
production at collider energies through our introduction of  the flavoring
factors.~\cite{GlobalFlavoring}   

 Let us begin by first examining  what we should  expect. Concentrate on  the triple-Pomeron vertex
$g(0)$ measured at high energies. Let us for the moment assume that it has also been meassured
reliably at low energies, and let us denote it as
$g^{low}(0)$. Our flavoring analysis  indicates that these two couplings are related by
$g_{\cal PPP}(0) \simeq e^{-({3\epsilon y_f\over 2})}g_{\cal PPP}^{low}(0).
$
With $\epsilon\simeq 0.08\sim  0.1$ and $y_f\simeq 8\sim 10$,
using the value $g_{\cal PPP}^{low}(0)=0.364\pm 0.025\>\> mb^{\half}$, we expect a value of
$0.12\sim 0.18\> mb^{\half}$ for $g_{\cal PPP}(0)$. Denoting the overall multiplicative constant for
our renormalized triple-Pomeron formula by
$
K\equiv {\beta^2_a(0)g_{\cal PPP}(0)\beta_b(0)/ 16\pi}.
$
With $\beta^2_p\simeq 16\> mb$, we therefore expect $K$ to lie between the range $.15\sim .25\>\>
mb^2$.

We begin testing our renormalized triple-Pomeron formula by first
determining the overall multiplicative constant $K$ by normalizing the
integrated $\sigma^{sd}$ to the measured CDF $\sqrt s=1800\> GeV$
value. With $\epsilon=0.1$, $\lambda_f=1$, this is done for a series
of values for $y_f=7,\>8,\>9,\>10$. We obtain respective values for
$K=.24,\> 0.21,\> 0.18,\> 0.15,$ consistent with our flavoring
expectation.  As a further check on the sensibility of these values
for the flavoring scales, we find for the ratio $\rho\equiv
\sigma^{sd}(546)/\sigma^{sd}(1800)$ the values $0.63, \> 0.65,\>
0.68,\> 0.72$ respectively. This should be compared with the CDF
result of 0.834.

Having shown that our renormalized triple-Pomeron formula does lead to
sensible predictions for $\sigma^{sd}$ at Tevatron, we can improve the
fit by enhancing the $PPR$-term as well as $RRP$-terms which can
become important. Instead of introducing a more involved
phenomenological analysis, we simulate the desired low energy effect
by having $\epsilon_o\simeq -0.06\sim -0.08$. A remarkably good fit
results with $\epsilon=0.08\sim 0.09$ and
$y_f=9$.~\cite{GlobalFlavoring} This is shown in
Figure~\ref{fig:diff_xs}. The ratio $\rho$ ranges from $0.78\sim
0.90$, which is quite reasonable. The prediction for $\sigma^{sd}$ at
LHC is $12.6\sim 14.8\> md$.

Our fit leads
to a triple-Pomeron coupling in the range of 
\be
g_{\cal PPP}(0)\simeq .12\sim .18 \>\>{mb}^{\half}, 
\ee
exactly as expected. Interestingly, the triple-Pomeron coupling quoted
in Ref. 3 ($g(0)=0.69 \>mb^{1\over 2}$) is actually a factor of 2
larger than the corresponding low energy value.~\cite{Tan1} Note that
this difference of a factor of 5 correlates almost precisely with the
flux renormalization factor $N(s)\simeq 5$ at Tevatron energies.

\section{ Final Remarks:}

In Ref. 1, a more elaborated treatment has been caried out where both
the flavoring and the final-state screening effects were
considered. We have shown, given Pomeron as a pole, the total Pomeron
contribution to a singly diffractive dissociation cross section can in
principle be expressed as 
\begin{eqnarray}
 {d\sigma \over
dtd\xi} & =& [S_i(s,t)][D_{a,{\cal P}}(\xi,t)] [\Sigma_{{\cal P} b} (M^2)],\\
 D_{a,{\cal P}} (\xi, t) & = &S_f(\xi,t)F_{{\cal P}/a} (\xi, t).  
\end{eqnarray}
\begin{itemize}
\item
 The first term, $S_i$, represents initial-state screening correction.  We have  demonstrated that,
with a Pomeron intercept greater than unity and with  a pole approximation for total and elastic
cross sections remaining valid, initial-state absorption cannot be large.  We therefore can justify 
setting 
$S_i\simeq 1$ at Tevatron energies.  

\item The first crucial step in our alternative resolution to the
Dino's paradox lies in properly treating the final-state screening, $S_f(\xi,t)$.
We  have explained in an expanding disk setting why a final-state screening can set in relatively 
early when compared with that for elastic and total cross sections.

\item
 We have stressed that the dynamics of a soft Pomeron in a
non-perturbative QCD scheme requires taking into account the effect of ``flavoring", the notion that
the effective degrees of freedom for Pomeron is suppressed at low energies. As a consequence, we
find that 
$ F_{{\cal P}/a} (\xi, t)=R^2(\xi^{-1}) F^{cl}_{{\cal P}/a} (\xi, t)$ and  $ \Sigma_{{\cal
P}b}(M^2)=R(M^2)\Sigma_{{\cal P}b}^{cl}(M^2)$ where $R$ is the ``flavoring" factor discussed in
this paper. 
\end{itemize}

It should be stressed that our discussion depends crucially on the
notion of soft Pomeron being a factorizable Regge pole. This notion
has always been controversial.  Introduced more than thirty years ago,
Pomeron was identified as the leading Regge trajectory with quantum
numbers of the vacuum with $\alpha(0)\simeq 1$ in order to account for
the near constancy of the low energy hadronic total cross
sections. However, as a Regge trajectory, it was unlike others which
can be identified by the particles they interpolate. With the advent
of QCD, the situation has improved, at least conceptually. Through
large-$N_c$ analyses and through other non-perturbative studies, it is
natural to expect Regge trajectories in QCD as manifestations of
``string-like" excitations for bound states and resonances of quarks
and gluons due to their long-range confining forces. Whereas ordinary
meson trajectories can be thought of as ``open strings" interpolating
$q\bar q$ bound states, Pomeron corresponds to a ``closed string"
configuration associated with glueballs. However, the difficulty of
identification, presummably due to strong mixing with multi-quark
states, has not helped the situation in practice. In a simplified
one-dimensional multiperipheral realization of large-N QCD, the
non-Abelian gauge nature nevertheless managed to re-emerge through its
topological structure.~\cite{LeeVeneziano}

 The observation of ``pole dominance" at collider energies has
 hastened the need to examine more closely various assumptions made
 for Regge hypothesis from a more fundamental viewpoint. It is our
 hope that by examining Dino's paradox carefully and by finding an
 alternative resolution to the problem without deviating drastically
 from accepted guiding principles for hadron dynamics, Pomeron can
 continued to be understood as a Regge pole in a non-perturbative QCD
 setting. The resolution for this paradox could therefore lead to a
 re-examination of other interesting questions from a firmer
 theoretical basis.  For instance, to be able to relate quantities
 such as the Pomeron intercept to non-perturbative physics of color
 confinement represents a theoretical challenge of great importance.

{\bf Acknowledgments:} I would like to thank K. Goulianos for first
getting me interested in this problem. I am also grateful to
P. Schlein for explaining to me details of their work.  Lastly, I
appreciate the help from K. Orginos for both numerical analysis and
the preparation for the figures. This work is supported in part by the
D.O.E.  Grant \#DE-FG02-91ER400688, Task A.

\newpage

\begin{figure}
\begin{center}
\includegraphics[scale=.65]{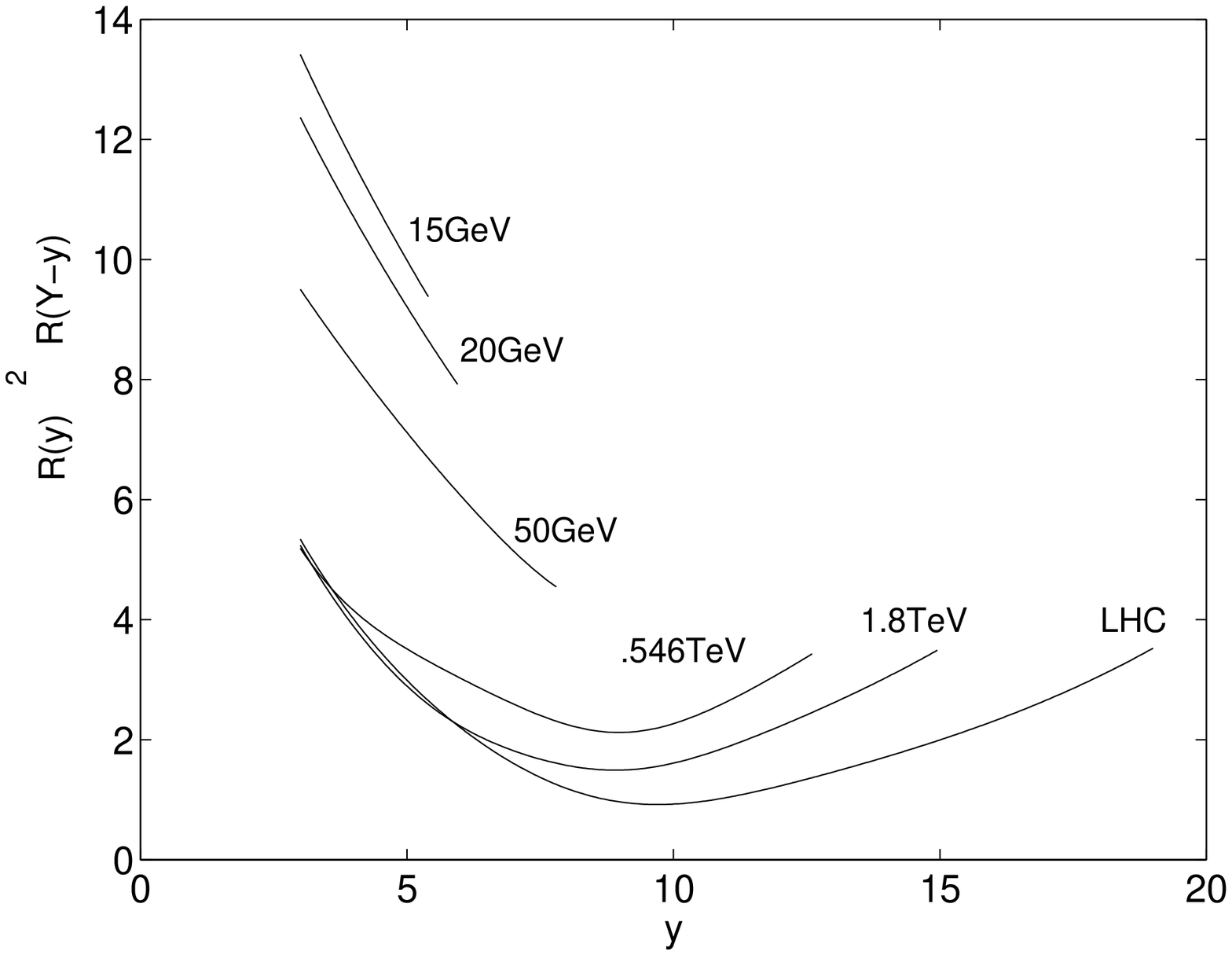} 
\end{center}
\vspace{.1cm }
\caption{Renormalization factor, $Z(\xi;s)\equiv R^2(\xi^{-1})R(M^2)$,
as a function of rapidity $y=\log \xi^{-1}$ for various fixed center
of mass energies.  These curves correspond to parameters used for the
solid line in Figure 2.}
\label{fig:flavoring} 
\end{figure}

\begin{figure}
\begin{center}
\includegraphics[scale=.65]{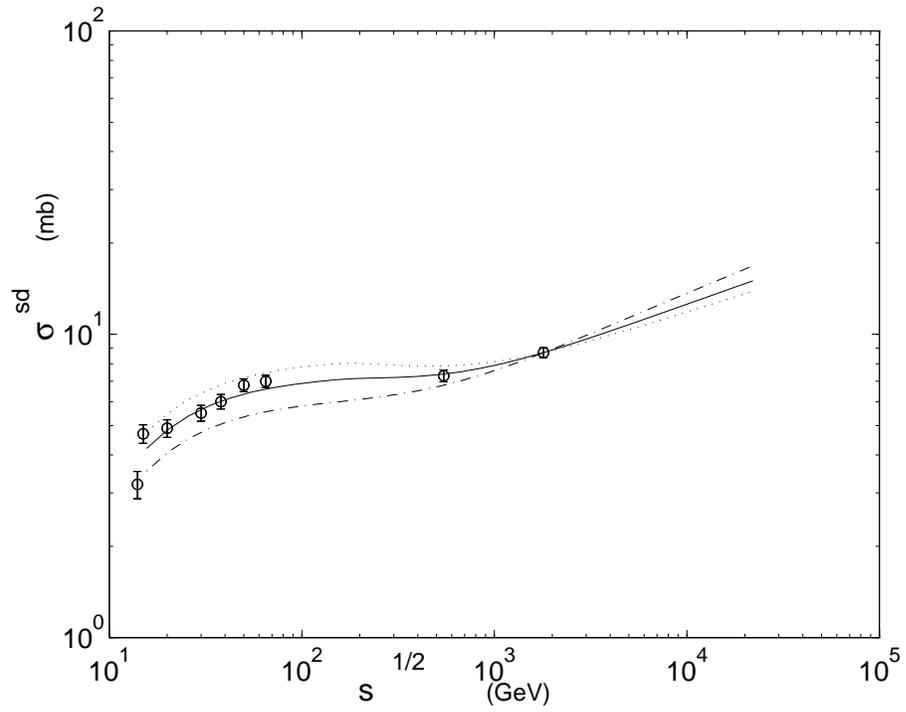} 
\end{center}
\vspace{.1cm }
\caption{Various fits to representative single diffraction
dissociation cross sections extracted from Ref. 3 from ISR to
Tevatron. The solid line and the dotted curve correspond to
$\epsilon=0.08$, $\epsilon_o=-0.07$, $\lambda_f=1$, $y_f=9$, with
small amount of final-state screening.  The dashed-dotted curve
corresponds to no screening.}
\label{fig:diff_xs} 
\end{figure}


\begin{thebibliography}{99}
\bibitem{Classical}{We use $s_0= 1$ $ GeV^2$ as the basic energy scale
through out this paper. These ``classical" expressions are: $F_{{\cal
P}/a}^{cl} (\xi, t)=(1/16 \pi)\beta_a(t)^2 (\xi^{-1})^{2\alpha_{\cal
P}(t) -1},$ and $\sigma_{{\cal P}b}^{cl}(M^2,t)=g(t)
(M^2)^{\alpha_{\cal P}(0)-1}\beta_b(0).$ The triple-Pomeron coupling
$g(t)$ can be in principle determined by data below $\sqrt s\leq 30\>
GeV$ where cross sections are relatively insensitive to the choice of
the Pomeron intercept value used.  }
\bibitem{GlobalFlavoring} By choosing $\epsilon^{old}< 0
$, it is possible to provide a global ``average" description mimicking
``secondary trajectory" contributions for various low energy
regions. In Ref. 5, acceptable estimates are $\epsilon_o\simeq
-0.11\sim -0.5$.
\end{thebibliography}
\end{document}